\documentclass[aps,prb,twocolumn,showpacs,superscriptaddress,amsfont,amsmath,amssymb,floatfix,10pt,reprint]{revtex4-2}
\usepackage{graphicx}  
\usepackage{dcolumn}   
\usepackage{bm}        
\usepackage{amssymb}   
\usepackage{framed}
\usepackage{amsmath}
\usepackage{multirow}
\usepackage{mathtools}
\usepackage{physics}
\usepackage{longtable}
\usepackage{hhline}
\usepackage{xcolor}
\definecolor{bluegreen}{rgb}{0,0.2,0.8}
\usepackage{subfigure,amsmath,verbatim,moreverb}
\usepackage{tabularx}
\usepackage{adjustbox}
\usepackage{lipsum}
\usepackage{longtable}
\usepackage{booktabs}
\usepackage{latexsym}
\usepackage{epsf}
\usepackage{float}
\usepackage{todonotes}
\usepackage{multibib}
\usepackage{booktabs}
\usepackage{array}
\usepackage{ragged2e}
\usepackage{tabularx}
\newcolumntype{Y}{>{\RaggedRight\arraybackslash}X}
\usepackage[colorlinks=true, linkcolor=blue, citecolor=blue, urlcolor=blue]{hyperref}
\usepackage{braket}
\usepackage{makeidx}
\makeindex
\usepackage{etoolbox}
\AtBeginEnvironment{align}{\setcounter{subeqn}{0}}
\newcounter{subeqn} %

\begin{document}

\thispagestyle{empty}

\title{Spin-orbit coupling and beyond in Chiral-Induced Spin Selectivity}

\author{Ruggero Sala}
\thanks{Equally contributed to this work.}%
\affiliation{Department of Chemistry \& INSTM, Università degli Studi di Pavia, Via Taramelli 12, 27100 Pavia, Italy.}
\affiliation{University School for Advanced Studies (IUSS), Piazza della Vittoria 15, 27100 Pavia, Italy.}
\author{Sushant Kumar Behera}
\email{sushantkumar.behera@unipv.it}
\thanks{Equally contributed to this work.}%
\affiliation{Department of Physics, Università degli Studi di Pavia, Via Bassi 6, 27100 Pavia, Italy.}
\author{Abhirup Roy Karmakar}
\affiliation{Department of Chemistry, Università degli Studi di Milano, Via Golgi 19, 20133 Milano, Italy.}
\author{Matteo Moioli}
\affiliation{Department of Physics, Università degli Studi di Pavia, Via Bassi 6, 27100 Pavia, Italy.}
\author{Rocco Martinazzo}
\email{rocco.martinazzo@unimi.it}
\affiliation{Department of Chemistry, Università degli Studi di Milano, Via Golgi 19, 20133 Milano, Italy.}
\author{Matteo Cococcioni}
\email{matteo.cococcioni@unipv.it}
\affiliation{Department of Physics, Università degli Studi di Pavia, Via Bassi 6, 27100 Pavia, Italy.}

\date{\today}

\begin{abstract}
Chiral-Induced Spin Selectivity (CISS) describes the emergence of spin-polarized electron transport in chiral systems without magnetic fields—a remarkable effect in light-element materials with weak intrinsic spin–orbit coupling (SOC). This mini-review analyzes the microscopic origins of CISS, highlighting how molecular chirality, local electric fields, and dynamic distortions enhance effective SOC and drive spin-dependent transport. We critically assess existing models in terms of their symmetry constraints, phenomenological assumptions, and compliance with Onsager reciprocity. Recent developments combining relativistic quantum mechanics and complete multipole representations reveal a direct link between chirality density and spin current pseudoscalars, suggesting a field-theoretic foundation for CISS. These insights could help positioning light-element chiral nanomaterials as tunable platforms for probing and engineering spin-selective phenomena at the nanoscale.

\end{abstract}

\maketitle

\section*{Introduction}
Chirality-Induced Spin Selectivity (CISS) is a remarkable phenomenon in which electron transport through chiral systems leads to spin polarization without the need for external magnetic fields \cite{Brian2024}. First observed in 1999 \cite{Kaushik1999, Naaman2012}, CISS has been reported in biomolecules like DNA \cite{Gohler2011}, helical and supramolecular structures \cite{Anup2017, Kulkarni2020}, and non-centrosymmetric materials \cite{Naaman2019}, with spin polarization approaching 100\% in some cases \cite{Haipeng2020, Al-Bustami2022}. 
Due to its broad phenomenology the CISS effect has attracted a significant interest due to its potential applications in a large variety of technologies, like spintronics \cite{Naaman2015}, quantum computing \cite{Wang2024}, redox chemistry \cite{Anupkumar2017}, and optoelectronics \cite{Long2020}. At its core, CISS arises from the interplay between electron spin and structural chirality, rooted in broken spatial inversion symmetry \cite{Bing2023}. However, the microscopic origin of this effect remains unresolved \cite{Pius2025}. While spin–orbit coupling (SOC) is often invoked as a key ingredient, many CISS-active systems consist mainly of light atoms, suggesting that conventional atomic SOC is insufficient. Alternative mechanisms include enhancement of SOC by chirality-induced electric fields \cite{Guinea2010, Abdiel2024}, geometric effects like curvature and torsion \cite{Levy2010, Binghai2024}, electron correlation effects \cite{Yang2025}, and spin-dependent scattering with chiral phonons or surfaces \cite{Fransson2023} to realize the CISS effect in nanomaterials \cite{Huang2025, Tabassum2024}. \\
\begin{figure*}
 \centering
 \includegraphics[height=9.5cm]{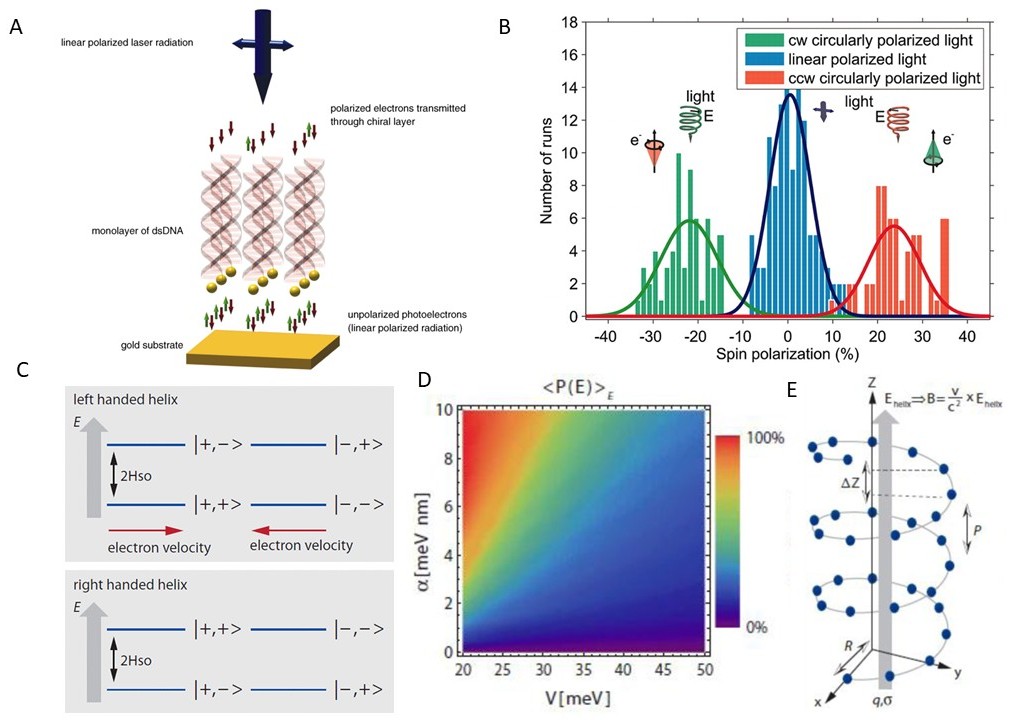}
 \caption{
 Selection of various phenomena that can be traced back to CISS. 
 (A) Schematic of a dsDNA monolayer as a spin filter: unpolarized electrons ejected from Au by linearly polarized light become spin-polarized antiparallel to their velocity, while non-transmitted electrons tunnel back to the substrate. (B) Spin polarization of photoelectrons from Au(111) with cw (green), ccw (red), and linear (blue) light ($-22\%$, $+22\%$, $0\%$, respectively); dsDNA/Au(111) transmits spin-polarized electrons. (C) Energy scheme of $|{\rm momentum, spin}\rangle$ states in a chiral potential; spin flips with helix handedness. (D) $\langle P(E)\rangle$ vs coupling $V$ and SOC $\alpha$: polarization increases for small $V$ and large $\alpha$. (E) Charge $q$ in spin state $\sigma$ moving along a helical charge distribution (pitch $p$, radius $R$, spacing $\Delta z$); the helical field $\vec{E}_{\rm helix}$ induces $\vec{B}$ affecting spin. 
 (A) and (B) reproduced from Ref. \citenum{Gohler2011} with permission. Copyright 2011, American Association for the Advancement of Science. (C), (D) and (E) reproduced from Ref. \citenum{Naaman2012} with permission. Copyright 2012, American Chemical Society. 
 }
 \label{figure1}
\end{figure*}
\begin{figure*}
 \centering
 \includegraphics[height=9.2cm]{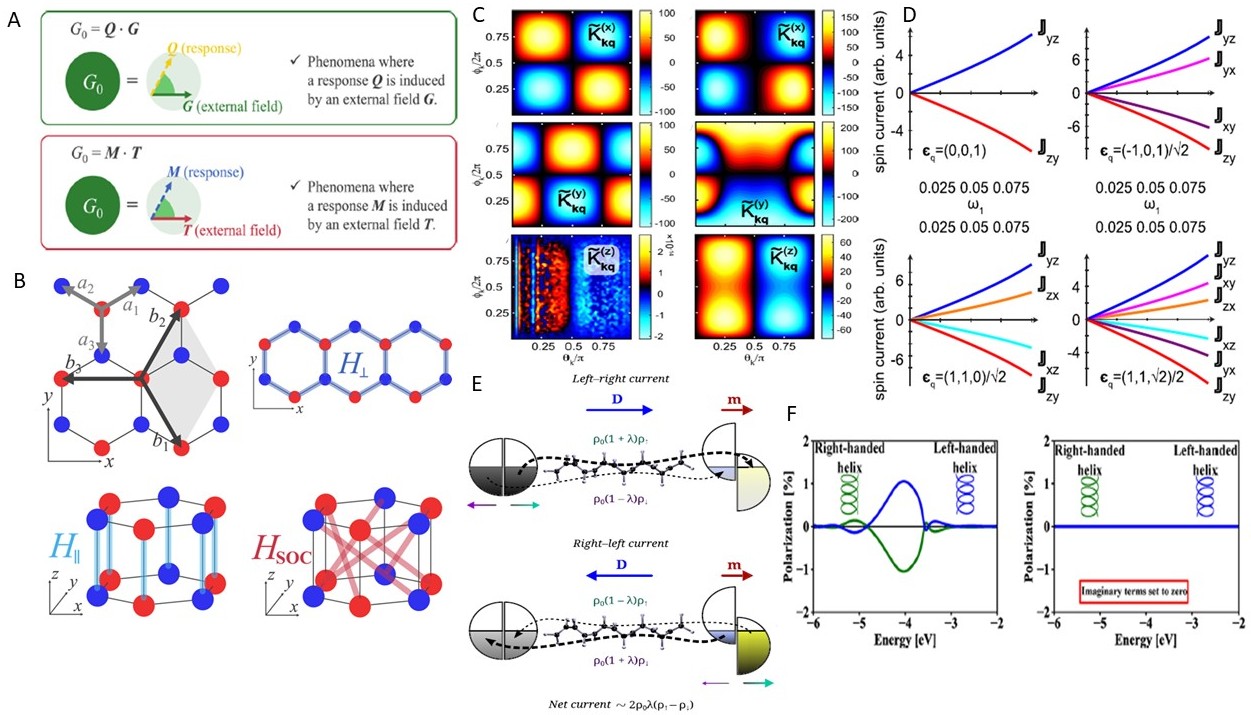}
 \caption{
 Schematic illustration of various mechanisms proposed as possible explanations of CISS. 
 (A) Schematic representations of the pseudoscalar quantity $G_0$, associated with electric polarization and current-induced magnetization, highlighting its symmetry properties. 
 (B) Tight-binding model on a honeycomb lattice including SOC-induced chiral hopping ($\chi = -1$). Reproduced from Ref.~\citenum{Hirakida2025} with permission. Copyright 2025, American Physical Society. (C) Examples of $\tilde{K}_{kq}$ (electron--phonon coupling kernel) 
 for achiral and chiral phonons, illustrating the symmetry differences in momentum space. (D) Spin--current tensor components as a function of the phonon mixing parameter $\omega_1$. Reproduced from Ref.~\citenum{Fransson2023} with permission. Copyright 2023, American Physical Society. (E) Schematic of spin-selective electron transport through a chiral molecule in the presence of SOC. Reproduced from Ref.~\citenum{Dalum2019} with permission. Copyright 2019, American Chemical Society. (F) Calculated spin polarization of the transmitted current: 
 negligible for linear molecules and on the order of $\sim$1\% for helical molecules, with the sign reversing upon changing handedness. The difference arises from the presence (helical case) or absence (linear case) of structural chirality and the associated SOC-induced phase terms; in the linear case, the chiral (imaginary) hopping components are set to zero. Reproduced from Ref.~\citenum{Zollner2020} with permission. Copyright 2020, American Chemical Society. 
 }
 \label{figure2}
\end{figure*}
Beyond the differences between the various 
experimental setups and definitions, ranging from asymmetric scattering 
to spin-selective transport and chiral interactions with magnetized surfaces \cite{Koyel2018}, all CISS manifestations reflect a coupling between spin and molecular handedness. 
The lack of a unifying theory has motivated several transport-based models, including Green’s function approaches and the Landauer–Büttiker formalism \cite{Buttiker1993, Luo2020}, which relate spin-resolved transmission to spin current and polarization \cite{Huisman2023, Pasquale2025}. 

Recent theoretical developments suggest that spin dynamics and spin current, not just charge transport, 
are central to understanding CISS 
\cite{Kaushik1999, Naaman2019}. Dynamical effects were recognized to be important also in Ref. \citenum{Vittmann2023}, where low-frequency vibrational modulations of SOC in helical models were shown to yield significant spin polarization, even in nominally single-channel systems where spin polarization is forbidden by time-reversal symmetry. Fig. \ref{figure1} illustrates some important aspects of CISS: the spin-filtering action of dsDNA on the electronic photocurrent from an Au surface (A); the dependence of the spin-polarization of photo-electrons from Au(111) in dependence of the polarization of the incident light (B); the energy scheme of electronic states in a chiral potential (C); the spin polarization of the current as a function of the applied voltage and of the SOC intensity (D); the charge q and spin $\sigma$ moving through an helical distribution (E). \\
A general theoretical framework to rationalize CISS can be derived from relativistic quantum mechanics, which being rooted in the Dirac equation provides a compelling foundation by inherently linking spin and chirality, suggesting that CISS may ultimately stem from fundamental quantum principles. New tools to describe spin–chirality interactions at the microscopic level are also provided by the multipole expansion of various relevant quantities involved in CISS which is proving a very powerful formalism to describe (and unify) the different manifestations of this phenomenon.   \cite{Hirakida2025}.  \\

In this mini-review, we explore how spin–orbit effects and spin dynamics contribute to the emergence of CISS, with an emphasis on theoretical models and symmetry-based interpretations. Our aim is to clarify the conceptual landscape and highlight promising directions for future research toward a unified theoretical framework. 

Some of the proposed theoretical framework and mechanisms to understand CISS and explain its complex phenomenology are illustrated in Fig. \ref{figure2}. These include: an example of a multipole representation of relevant quantities (A); a TB model with SOC-induced chiral hoppings (B); examples of the influence of chiral phonons on the electronic current (C and D); a scheme to understand the role of SOC on inducing spin-selective transport in single chiral molecules (E and F). 

\section*{Theoretical Perspectives on CISS}

Understanding the Chiral-Induced Spin Selectivity (CISS) effect from a theoretical standpoint requires confronting several fundamental constraints. In particular, two major theoretical limitations were highlighted by Evers \textit{et al.}~\cite{Evars2022}. 

The first limitation concerns strictly one-dimensional transport channels—of particular relevance for molecular wires. In practice,
however, it is typically circumvented by multichannel transport, non-equilibrium conditions, dynamical spin–orbit coupling, as
well as dephasing and inelastic processes that unavoidably occur in realistic devices. The constraint originates from the combined
action of unitarity and time-reversal symmetry, which forbids spin polarization at the drain of a two-terminal device supporting
only a single orbital mode (i.e., a single Kramers pair). The proof is straightforward, and we sketch it here for completeness.
Consider an unpolarized current injected from the left lead. In the single-channel case, unitarity of the scattering operator $S$ implies that 
the spin polarization along an arbitrary quantization axis $Z$ at the drain can be written as $P_Z = (R_{\uparrow\uparrow}-R_{\downarrow\downarrow}) + (R_{\uparrow\downarrow}-R_{\downarrow\uparrow})$  where $R_{\sigma'\sigma}$ are spin-resolved reflection probabilities for electrons injected from the right lead. That is, $R_{\sigma'\sigma} = \left| \langle f\sigma' R|S|f\sigma R\rangle \right|^2$, where $S$ is the scattering operator and $f$ denotes the single orbital channel available in the right electrode ($R$). Under time reversal $\mathcal{T}$, the scattering operator transforms as $\mathcal{T} S \mathcal{T}^{-1} = \bar{S}^{\dagger}$
where the bar denotes the scattering operator of the time-reversed Hamiltonian. 
The time-reversal operator acts on channel states by flipping spin,
$\mathcal{T} \ket{f\tau R} = e^{i\varphi_\tau} \ket{f\bar{\tau}R}$ with $\bar{\tau}=-\tau$ and $\varphi_\sigma$ arbitrary, though $e^{i(\varphi_\uparrow+\varphi_\downarrow)}=-1$ due to $\mathcal{T}^2=-1$.  For a time-reversal invariant system $\bar{S}\equiv S$ and one finds $R_{\uparrow\uparrow}=R_{\downarrow\downarrow}$ and $R_{\uparrow\downarrow}=R_{\downarrow\uparrow}=0$, which together enforce $P_Z=0$. Thus, spin polarization is strictly forbidden in a phase-coherent, two-terminal device with a single-orbital-channel at the drain.\\

This “no-go theorem” is closely related to a theorem due to Bardarson (Ref.~\citenum{Bardarson_2008}), which has sometimes been misinterpreted as proving that spin polarization is impossible in any two-terminal time-reversal-invariant system. Bardarson showed that in such systems the transmission eigenvalues occur in degenerate pairs. This follows from a simple observation: in an appropriate Kramers-paired basis, the scattering matrix can be brought to an antisymmetric form, $S^{t}=-S$. The skew-Takagi decomposition of antisymmetric matrices then implies that the reflection eigenvalues are doubly degenerate and that the corresponding eigenvectors are related to each other. By unitarity, the same must hold for the transmission eigenvalues, apparently preventing spin-filtering effects. However, Bardarson’s degeneracy concerns eigenvalues, not the spin structure of the corresponding eigenvectors. 
The degenerate transmission eigenchannels enforced by time-reversal symmetry are not related to each other by time reversal itself, as would be required for a genuine Kramers pair, rather by the reflection block of S. The single-orbital-channel case is exceptional because (with Bardarson’s choice of channel basis) $S\propto \sigma_y$ in each orbital subspace. In this situation, if the right electrode supports only a single orbital channel, its (degenerate) \emph{reflection} eigenchannels carry opposite spin polarization (due to the action of the reflection $S\propto \sigma_y$) and, by unitarity, the same must hold for the left-to-right transmission channels. That is, the theorem only addresses the degeneracy of the transmission eigenchannels, but time-reversal symmetry alone does not forbid spin polarization in multichannel two-terminal devices (see, e.g., Ref.~\citenum{Utsumi2020}).\\



The second, more critical limitation relates to Onsager’s reciprocal relations and concerns the vanishing of the zero-bias magnetoconductance,
i.e., $\Delta G(\mathbf{B},V) = G(\mathbf{B},V) - G(-\mathbf{B},V)\rightarrow 0 $ as $V\rightarrow0$, a condition that is typically violated in CISS experiments. A typical CISS transport experiment involves a CISS spin-valve, where a chiral media is sandwiched between a normal metal and a magnetic layer, and the magnetization $\mathbf{M}$ of the latter can be switched with external magnetic fields. The current-voltage ($I-V$) curves and the differential conductance $G(\mathbf{M},V)=dI/dV$ are then measured for opposite values of the magnetization, defining the above $\Delta G$ (with $\mathbf{M}$ playing the role of $\mathbf{B}$). This magnetoconductance is used as a proxy for spin polarization of the current, and Onsager’s statement is sometimes mistakenly interpreted as implying the absence of (transmitted) spin polarization in equilibrium. However, $\Delta G$, although sensitive to spin polarization, is only loosely - and at best heuristically - related to it. Even more problematic is the widespread habit of quantifying it as $\textrm{MCR}(\textbf{M})=\Delta G (\mathbf{M})/(G(\mathbf{M})+G(\mathbf{-M}))$ rather than reporting the directly measured magnetoconductance \cite{Liu2023}. 
This magnetoconductance ratio (or the 
\textcolor{blue}{equivalent} magnetoresistance ratio, MRR$=-$MCR) is often quoted as the "measured" spin polarization~\cite{Das2025}, even though no direct measurement of the spin state of the current is actually performed. The actual situation is rather intricate. First, when magnetic fields $\mathbf{B}$ (rather than ferromagnetic electrodes $\mathbf{M}$) are employed, various orbital effects, such as Aharonov-Bohm phases, weak localization or antilocalization corrections, and Landau quantization, can modify the conductance in ways unrelated to spin. Second, even barring these "orbital" effects, there is no direct connection between the spin-state of the current and the 
magnetoresistance.
In order to 
relate MCR 
to the spin polarization of the chiral material one often invokes 
Julliere's 
tunneling model \cite{Julliere1975}. 
used for magnetic tunnel junctions 
However, this interpretation is inconsistent with experimental observations \textcolor{blue}{of CISS} showing that 
MCR can exceed the intrinsic spin polarization of the ferromagnetic electrode (typically $\sim30\%$), which should in principle represent its upper bound.\\
On the other hand, 
spin polarization of the current is 
generally finite even in equilibrium ($V = 0$) and in the absence of any magnetic field, 
\emph{without} contradicting time-reversal symmetry. 
It may be very small, but generally non-vanishing, unless specific spatial symmetries forbid it \cite{Dednam2023}. 
Spin polarization describes the spin state of the \emph{transmitted} current and is therefore an intrinsically dynamical quantity, which is not simply inverted upon time reversal. 
That is, even in an equilibrium state, spin polarization refers, by definition, to the spin state of the current collected at a drain terminal when a (typically unpolarized) current is \emph{injected} from a source terminal (more details in the Appendix) 
\cite{nikolic2005}.\\
To summarize, magnetoconductance  is not directly linked to the spin polarization of the current, and breaking Onsager reciprocity is not required for spin polarization to occur. 
CISS magnetoresistance effects and the intrinsic spin-polarizing response of the chiral medium are (likely) related but physically distinct phenomena that must be treated separately. 
A violation of Onsager reciprocity, however, serves as a clear indication that the system is driven out of true equilibrium. The accumulation of chirality-dependent charge due to ordinary electric-magnetochiral anisotropy 
\cite{Rikken2001}
could then account for the observed symmetry upon magnetization reversal and simultaneous change in handedness \cite{Zhao2025}. 
Therefore, theoretical models aiming to describe CISS 
phenomenology must involve either a symmetry breaking or consideration of non-linear transport regimes. 
This theoretical backdrop sets the stage for the various modeling efforts aimed at explaining CISS.

\subsection*{Current Theoretical Models}
The majority of theoretical approaches to the CISS focuses on spin-dependent transport through chiral molecules (see Fig. \ref{figure2} \textcolor{blue}{for some examples}). 
Most transport-based theoretical models rely on the Landauer–Büttiker formalism \cite{Datta1995} 
in combination with Green’s function methods, which allow one to calculate spin-dependent transmission coefficients from the system’s Hamiltonian in the phase-coherent regime. 
These calculations typically define a central region (the chiral molecule) connected to metallic leads, from which charge currents can be computed under steady-state conditions, using Green's function-based scattering theory and, occasionally, DFT Hamiltonians. Theoretical strategies to circumvent the limitations of phase-coherent scattering often involve introducing phase-breaking and inelastic scattering via Büttiker probes or adding phenomenological leakage terms (shown in Fig. \ref{figure2}). 
Although effective, these additions are frequently \textit{ad hoc} and lack a clear physical justification, highlighting the need for more rigorous microscopic foundations. 
Furthermore, static models may be intrinsically inadequate to capture the essence of CISS. For instance, a dynamical SOC arising from low-frequency vibrations may significantly contribute to its onset by generating a finite spin polarization and/or introducing inelastic and phase-breaking effects\cite{Vittmann2023}. And, at a more basic level, even without invoking electron–phonon interactions, one should always account for the various molecular conformations potentially involved in the scattering process (arising either from atomic vibrations or, more generally, from the intrinsic flexibility of organic molecules 
contacted between electrodes). 
%
Two common approaches are used to construct the Hamiltonians in CISS simulations: \textit{ab initio} density functional theory (DFT) and parameterized tight-binding (TB) models. DFT-based NEGF calculations, though more physically grounded, are computationally demanding and less flexible in terms of device geometry, yet they can naturally incorporate key screening effects in non-equilibrium situations, thereby effectively breaking time-reversal invariance of the self-consistent Hamiltonian. \\
In contrast, TB models allow greater control over specific parameters, making them popular for exploring qualitative aspects of CISS. Both approaches can, as mentioned above, include virtual probes through phenomenological self-energy terms. Alternative frameworks, such as wavefunction scattering-matrix formalisms, have also been proposed, though they are less widely adopted 
due to their inability in describing screening effects. 
Dalum and Hedegård~\cite{Dalum2019} (DH) introduced a perturbative NEGF model that captures CISS responses 
through an auxiliary axial vector 
capable to describe spin polarization of the scattered electrons. 
However, the results 
of Ref. \citenum{Dalum2019} 
were based on idealized TB models of twisted polyacetylene, and a more realistic, \textit{ab initio} implementation of the vector formalism remains a future goal. In fact, phenomenological ingredients are required for a sizeable magnetization to arise in equilibrium when non-magnetic electrodes are used. Breaking time-reversal symmetry by such means enhances the magnetization and allows it to emerge without the need for ferromagnetic electrodes.\\
Zollner \textit{et al.}~\cite{Zollner2020} 
explored the role of SOC in NEGF transport by showing that SOC introduces "imaginary" (spin-dependent) terms in the Hamiltonian — that is, imaginary matrices when operators are expressed in a time-reversal-invariant (spatial) basis — which are essential for maintaining Hermiticity rather than indicating broken time-reversal symmetry. These terms are sufficient for non-zero spin polarization to coexist with inversion symmetry, suggesting that chirality is not even a prerequisite for spin polarization. 
A careful symmetry analysis~\cite{Dednam2023} further reveals that the presence of a longitudinal mirror plane in the transport device, as may occur for certain achiral molecules and contact configurations, imposes no constraints on the transverse spin
polarization but forces the longitudinal component to vanish. The transverse component, however, also vanishes if the molecular
junction possesses a longitudinal symmetry axis~\cite{Dednam2023}. 
A similar cancellation is expected when transport is dominated by the intrinsic molecular properties of the junction rather than by
the interface,
and the molecule is either free to rotate about that axis or can exist in multiple, energetically quasi-equivalent, rotated configurations. 
These individual conformations are generally asymmetric and can
produce spin polarization, but their contributions cancel once
configurational averaging is taken into account, leaving room only for a longitudinal spin-polarization. 
Structural chirality becomes crucial precisely at this stage: in
achiral molecules a mirror plane exists which enforces
complete cancellation of the polarization of the ensemble; for chiral molecules,
on the other hand, the accessible configurational space is
intrinsically symmetry-broken and lacks such mirror symmetry,
thereby allowing a net (longitudinal) spin polarization to emerge.\\

It is in this sense that SOC-induced terms embody a manifestation of true chirality, as defined by Barron \cite{Barron1986}, where time-reversal symmetry is preserved while parity is broken. Yet, the apparent $PT$ symmetry observed in many CISS transport experiments points to deeper, symmetry-based mechanisms underlying spin selectivity, false equilibrium, and symmetry breaking - all ultimately rooted in the relativistic structure of space–time and its transformations. 
The structure of the fundamental homogeneous Lorentz group, when extended to complex transformations, naturally accommodates $PT$ as a fundamental symmetry of physical systems. 
Within this broader perspective, phenomena such as the non-Hermitian skin effect, recently invoked in the explanation of some CISS-related effects\cite{Zhao2025}, 
emerge naturally as manifestations of complex symmetry operations, further linking CISS phenomenology to non-hermitian physics. 
This perspective further aligns with more advanced theoretical frameworks, such as the multipole formalism recently introduced in the literature~\cite{Hirakida2025}.
On the other hand, the role of time-reversal symmetry in CISS remains an active topic of debate per se \cite{fransson23,fransson25,Amnon2025,Barron1986,barron20}. While some interpretations emphasize the need of T-symmetry breaking under non-equilibrium or magnetic boundary conditions, others support the idea that spin selectivity, as an intrinsic property of the systems, does not require a 
violation of T symmetry \cite{Utsumi2020,Nakajima2023,Yang2020}, as the latter is already broken when a current flows through the system, as already discussed above. 
\subsection*{New Directions for CISS Theory}
Theoretical models of CISS are increasingly focusing on reconciling the discrepancy between 
observed spin polarization and the relatively weak intrinsic spin-orbit coupling (SOC) expected in organic molecules composed of light atoms. In tight-binding effective Hamiltonians, the SOC values required to reproduce the experimental spin polarization often exceed several tens of meV - orders of magnitude higher than the atomic SOC that can be expected in the chiral molecules typically investigated. Early approaches that considered only electronic degrees of freedom, without coupling to vibrational or environmental modes, typically failed to account for the observed magnitude of the effect. There have been  attempts to attribute the enhanced SOC to the metallic leads as, for example, in Ref \citenum{Alwan2021}, thereby invoking orbital-to-spin angular momentum conversion or spin-transfer torque at the interfaces. However, these explanations of the CISS effect have encountered mixed success. For instance, while Adhikari \textit{et al.}~\cite{Adhikari2023} observed such an effect in a low-temperature spin valve, Kettner \textit{et al.}~\cite{Kettner2018} found that spin polarization in helicene monolayers remains largely invariant across different leads (Cu, Ag, Au), despite the huge differences that SOC has on the electronic states involved, undermining the lead-dominated SOC hypothesis. \\
To address these limitations, several theoretical efforts have proposed alternative mechanisms or incorporated additional bosonic and fermionic degrees of freedom within DFT or tight-binding frameworks. In Ref. \citenum{Dianat2020} authors propose that electronic exchange couplings can significantly influence the interaction between chiral molecules and magnetic substrates. In a different proposal by Fransson and coworkers, ~\cite{Fransson2023,Zhang2023} electron-electron interactions lead to spin-selective exchange fields, yielding finite spin polarization even in the absence of large atomic SOC. Similarly, Fay and Limmer described spin-selective recombination in donor-bridge-acceptor systems via incoherent hopping pathways, with spin selectivity emerging from a non-adiabatic regime of spin-orbit interactions~\cite{Fay2023}. Electron-phonon couplings have also been shown to play a significant role: Fransson demonstrated that the spin-dependent electron-vibration coupling, 
in presence of finite SOC,
introduces an exchange splitting between spin channels, yielding polarization on the order of tens of percent using realistic parameters~\cite{Fransson2020}. Polaron-based mechanisms have also been proposed as a route to enhanced spin selectivity in chiral systems ~\cite{Zhang2020, Oiwa2022}. Zhang \textit{et al.} further supported these findings with a polaron transport model, where strong coupling between electronic motion and lattice vibrations in a helical scaffold produced comparable levels of spin selectivity~\cite{Zhang2023}. 
Vittman et al. \cite{Vittmann2023} introduced the concept of dynamic SOC and demonstrated the critical role of low-frequency vibrations in generating a finite spin polarization in an otherwise ineffective single-channel system.  
In these approaches, strong electron–phonon couplings leads to polaron formation, where lattice distortions dynamically accompany charge transport. The interplay between chirality, spin–orbit coupling, and polaronic dressing can significantly modify spin-dependent transport properties, suggesting that vibronic correlations may amplify CISS beyond purely electronic models.\\
On the other hand, some recent work\cite{hattori26} has pointed out the possibility that certain electronic orders can actually exhibit chirality-related properties even in non-chiral structures and without atomic displacement, thus stressing the central role of a \emph{spatial} symmetry breaking in the electronic subsystem.\\
Beyond tight-binding models~\cite{Kato2025}, group-theoretical approaches~\cite{Hayami2024,Hirakida2025} have offered insights into symmetry constraints governing CISS. Dednam \textit{et al.}~\cite{Dednam2023} showed that spin polarization can emerge from the chiral configuration of the entire device - including the relative orientation of leads - rather than the molecular structure alone. 
In such configurations, broken spatial symmetries alter the elements of the scattering matrix, enabling spin- and direction-dependent conductance even under equilibrium and coherent transport conditions~\cite{Liang2025}. As already emphasized above, finite spin polarization is generally expected, unless special spatial symmetries enforce cancellations, provided spin-orbit interactions are present. Such exact symmetry conditions are rare in practice, as the asymmetry of the molecular configuration or the electrode usually breaks them. 
In an alternative vibronic framework, Kato \textit{et al.}~\cite{Kato2025} proposed a model in which pseudo-Jahn-Teller coupling between nuclear angular momentum and electronic states in chiral molecules yields intrinsically asymmetric spin filtering. The spin selectivity arises solely from molecular chirality and vibronic coupling, achieving higher spin polarization than bare SOC mechanisms. However, this model does not account for spin-flip processes, which may contribute to the net spin polarization observed experimentally.
These directions suggest that the large effective SOC inferred from experiments may arise from emergent interactions or cooperative effects rather than purely atomic contributions~\cite{Behera2024}.

Moreover, emergent spin–momentum coupling grounded in quantum geometry and topological constraints are gaining traction. These are terms that emerge from boundary conditions, dynamical correlations, or non-Hermitian effects, even when explicit SOC terms are absent in the Hamiltonian. The distinction from conventional SOC, however, can be somewhat blurred. For instance, boundary conditions typically replace confining potentials, and thus the associated electric fields. It is precisely the resulting non-atomic (or structural) spin–orbit coupling associated with these fields that ultimately arises \cite{gao2026}. 
While current models remain largely phenomenological, their increasing sophistication provides a framework for future \textit{ab initio} studies, which may ultimately unravel the microscopic origins of the CISS effect~\cite{Pius2025}. Despite significant progress, a comprehensive microscopic theory of the CISS effect remains elusive. Recent perspectives emphasize that conventional electron-transfer theories may miss key ingredients such as electron–electron interactions, polaronic effects, and dissipation, suggesting that a full many-body description beyond standard approximations may be required to explain spin-dependent transport in chiral systems \cite{Naaman2026}.

\subsection*{Origin of large effective SOC in light-atom molecule}
\begin{figure*}
 \centering
 \includegraphics[height=5.5cm]{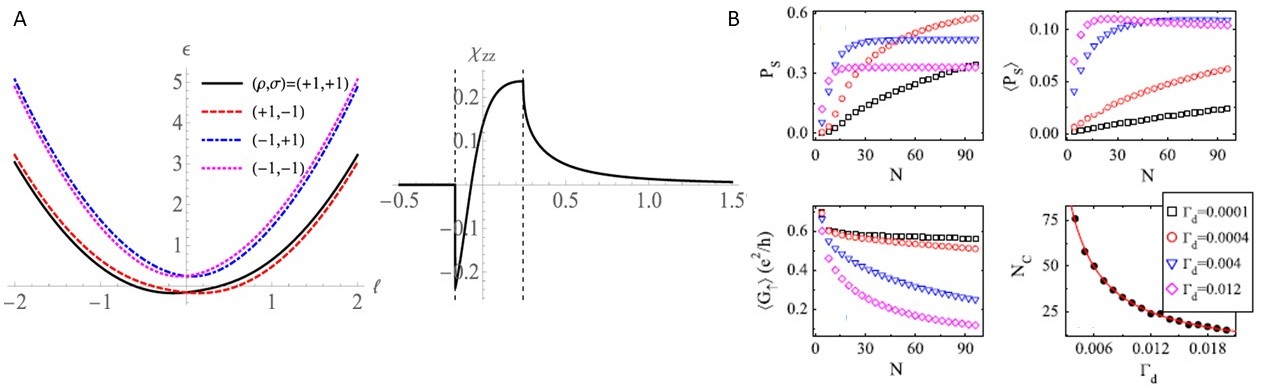}
 \caption{
 Spin response and transport in coupled-helix models and dephasing regimes.
 (A) Band structure of the coupled-helix model. The black (solid) and red (dashed) curves denote the two lower bands ($\rho=+1$), while the blue (dot-dashed) and magenta (dotted) curves denote the two upper bands ($\rho=-1$). Here, dimensionless Edelstein coefficient $\chi_{zz}$ for inter-helix coupling $\lambda=0.2$ as a function of chemical potential $\mu$. Dotted vertical lines indicate the band edges. Reproduced from Ref. \citenum{Shitade2020} with permission. Copyright 2020, Institute of Physics. 
 (B) Length dependence of spin polarization $P_s$ at $E=0.486$, $\langle P_s \rangle$, and conductance $\langle G^\uparrow \rangle$ with dephasing parameter $\tau_d$. Critical length $N_c$ vs. $\Gamma_d$, extracted from $\langle P_s \rangle$–$N$. The solid line shows the fit $N_c \propto \Gamma_d^{-1}$. Here, the legends correspond to other three panels. Reproduced from Ref. \citenum{Guo2012} with permission. Copyright 2012, American Physical Society.
 }
 \label{figure3}
\end{figure*}

As mentioned above, 
a persistent challenge in the theoretical descriptions of the CISS effect is reconciling the large spin polarization observed in chiral molecular systems with the intrinsically weak atomic SOC characteristic of light-element organic compounds \cite{Meng2024}. The canonical atomic SOC, typically treated as a relativistic correction to the nonrelativistic Schrödinger equation, is highly localized near the nuclei and scales with the fourth power of the atomic number, \( Z \), rendering it negligible in systems composed primarily of elements such as C, H, N, and O \cite{Malli2013, Eliel2008}. 
In addition, the effective SOC experienced by a transport channel is energy- and orbital-dependent, which can make the effects of this coupling negligible even in presence of heavy species. For example, in Au the conduction channel at the Fermi level is dominated by \textit{s}-electrons and therefore it is almost entirely insensitive to SOC 
\cite{Geyer2020}. Despite this, numerical agreement with experimental spin polarization often requires effective SOC values ranging from several meV up to hundreds of meV - orders of magnitude larger than what atomic considerations alone would predict. Understanding the origin of this enhanced effective SOC is thus essential for any physically consistent theory of CISS. In this context, however, we emphasize a point that appears to be largely overlooked in the literature: SOC is necessary to generate spin polarization, but at the same time SOC renders spin no longer a good quantum number. Thus, SOC must be present, yet not overwhelmingly strong—it should be present only to the extent required to enable spin–dependent effects ("just enough" or “\emph{quanto basta}”, q.b.). This may help explain the remarkable efficiency of light-element systems in producing spin polarization.
\\
One line of inquiry considers the emergence of SOCs from features of electronic structures beyond the atomic scale. Guo \textit{et al.},~\cite{Guo2012} in a tight-binding framework, modeled SOC in double-stranded DNA as arising from the gradient of the electrostatic potential at the helical boundary, rather than near atomic cores. The SOC term was taken in the general form as: \\ 
\begin{equation}
H_{\mathrm{SO}} = \frac{1}{4 m^2 c^2} \nabla V \cdot (\boldsymbol{\sigma} \times \mathbf{p}),
\label{Eqn2}
\end{equation} \\
where \( \boldsymbol{\sigma} \) is the vector of Pauli matrices, \( \mathbf{p} \) is the momentum operator, and \( \nabla V \) is the gradient of the electrostatic potential. This boundary-induced SOC, when combined with a two-channel transport model and local dephasing mechanisms (simulated \textit{via} virtual probes), yielded spin-polarized transmission consistent with experimental trends (see also Fig. \ref{figure3} B). Although this SOC was estimated to be on the order of a few meV, the model's effectiveness was attributed more to the inclusion of dephasing than the SOC magnitude itself. Importantly, the generality of boundary-induced SOC has not yet been systematically established for non-helical or more disordered chiral systems.\\ 
Additional studies have explored SOC contributions from circulating electronic currents in ring-like or helical molecular architectures, yielding similarly modest SOC values. While insufficient in isolation, these current-induced SOC terms become more relevant when embedded in spin-current or current-density functional theory (CDFT) frameworks \cite{Desmarais2024, Pittalis2017}, where current-dependent exchange-correlation potentials can introduce nontrivial spin-orbit effects \cite{Zollner2025}. These effects scale with the number of unpaired electrons and may thus be significant in open-shell systems. Furthermore, within this context, spin polarization may be understood as a dynamic redistribution of total angular momentum in response to chiral perturbations, a viewpoint advanced by Wolf \textit{et al.}~\cite{Wolf2022} and consistent with the conserved angular momentum current formalism proposed by An \textit{et al.}~\cite{An2012} Such perspectives support an orbital-to-spin angular momentum conversion mechanism and motivate further development of \textit{ab initio} current-dependent theoretical treatments tailored to chiral systems \cite{Bencheikh2003, Go2022, Jakub2017}. \\
From a symmetry-based standpoint, the complete multipole expansion formalism~\cite{Hayami2024} provides a rigorous method to derive asymmetric SOC terms (\textit{e.g.}, Rashba- and Dresselhaus-like interactions) even in non-periodic molecular systems. Moreover, this aspect connects chiral magnetic ordering to symmetry-breaking SOC terms. Foundational developments of symmetry-based multipole classifications~\cite{Kishine2022, Kusunose2024} provide a systematic framework to describe 
PT–odd responses, offering a rigorous language to classify chiral electronic and magnetic structures relevant to CISS. These terms arise naturally in low-symmetry environments and can, in principle, generate sizable spin-splitting and direction-dependent transport. In certain chiral crystals such as CrNb$_3$S$_6$, the observed large CISS signal (e.g., the generation of a sizeable bulk magnetization through the flow of an electric current along the principal axis of the crystal) has been attributed to antisymmetric SOC resulting from the crystal’s point-group symmetry \cite{Akito2020,Yoji2020}. However, the quantitative assessment of such terms in organic chiral molecules remains underexplored, and the role of vibronic coupling and dynamic disorder must be considered in realistic systems. 

A more fundamental approach to the SOC origin involves relativistic field theory. Shitade and Minamitani~\cite{Shitade2020} derived a geometric SOC by constraining the Dirac equation to a one-dimensional (1D) curved trajectory embedded in three-dimensional space and then taking its nonrelativistic limit (some results are shown in Fig. \ref{figure3}). The resulting SOC term scales as \( 1/m \), instead of the conventional \( 1/m^2 \),
leading to an effective SOC of approximately 160~meV, sufficient to account for observed spin polarization in DNA-like helical geometries \cite{Gohler2011, Guo2012} (see Fig. \ref{figure3} B). Intriguingly, this result depends sensitively on the order of the limits taken (geometric constraint versus non-relativistic approximation), indicating a non-trivial interplay between curvature and relativistic dynamics \cite{Abdiel2024}. In a related but distinct approach, Geyer \textit{et al.}~\cite{Geyer2020} derived a semi-relativistic Hamiltonian starting from the Pauli equation under strong radial confinement along a helical coordinate. Their model yielded spin polarization using SOC values in the 5~meV range, attributed to emergent non-Abelian gauge fields and Zeeman-like terms originating from radial scalar potentials. The disparity between the magnitudes of effective SOC in these models suggests that distinct physical mechanisms may be operative and highlights the need for a unified relativistic framework to compare and classify different spin-polarizing contributions in chiral systems. \\
Collectively, these diverse approaches suggest that the effective SOC in CISS-active light-atom systems is not simply an atomic property but an emerging consequence of the geometry of the system, electronic correlations, dynamic couplings, and relativistic corrections. The current landscape points to several promising directions for future exploration, including: (i) development of \textit{ab initio} current-dependent DFT methods capable of capturing orbital-spin conversion, (ii) multipole expansion-based derivations of SOC in realistic molecular geometries, and (iii) field-theoretic treatments incorporating geometry-induced SOC beyond the standard Pauli-Schr\"{o}dinger 
formalism. 
Although many of these methods are still in active development, they provide the theoretical infrastructure necessary to ground CISS in a quantitatively rigorous and microscopically meaningful way.

\subsection*{Minimal SOC Assumptions: Competing Theories}

An alternative theoretical perspective that has gained traction in the study of CISS explores mechanisms that operate with minimal SOC. These models invoke topological properties of electronic states, geometric (Berry) phases, non-adiabatic dynamics, and quantum geometry to explain spin polarization without relying on large atomic SOC values typical of heavy elements. 
This paradigm is particularly appealing for light-atom organic systems, where experimental evidence of significant spin polarization contrasts with the weak intrinsic SOC expected from atomic considerations. \\ 
One avenue involves the topological nature of electronic states. For example, Liu \textit{et al.}~\cite{Liu2022} proposed that chiral molecules such as DNA may exhibit nontrivial topological features in their electronic structure, potentially giving rise to spin-selective transport \cite{Gohler2011}. While this model is conceptually rich, it relies on a putative band structure that may not be applicable to the discrete energy levels of finite organic molecules used in CISS experiments. A related and more physically grounded insight comes from crystalline systems: in materials with broken spatial symmetries and magnetic textures, the interplay between spin and orbital degrees of freedom can lead to Hall-like responses and spin polarization even in the absence of strong SOC~\cite{Zhang2020_1}. For instance, CoNb$_3$S$_6$, an antiferromagnetic semimetal, hosts chiral Dirac fermions and shows an anomalous Hall effect despite weak ferromagnetic exchange \cite{Nirmal2018}. Although not yet tested for CISS, such materials may provide fertile ground for experimentally verifying spin-selective transport induced by chirality and magnetic structure. Along similar lines, chirality-induced spin polarization has recently been predicted in twisted transition metal dichalcogenides, where Moiré geometry and broken symmetries give rise to spin-selective responses even without invoking large atomic SOC~\cite{Menichetti2025, Menichetti2025_2}. These ideas are further supported by the complete multipole representation formalism~\cite{Hayami2024}, which connects chiral magnetic ordering to symmetry-breaking SOC terms, and by spin-current density functional theory (spin-current DFT), which has proven successful in modeling topological phases such as the quantum spin Hall effect~\cite{Desmarais2024}. \\
In parallel, concepts from geometric phase theory have been employed to understand spin polarization through nonadiabatic dynamics. The Berry phase and related quantities, the Berry curvature, the Berry connection, and antisymmetric geometric friction offer a unified language for phenomena as diverse as orbital magnetization~\cite{Berard2006, Berry2015}, polarization~\cite{Resta2000}, and various Hall effects~\cite{Guinea2010}. These geometric constructs have also been linked to molecular vibronic processes such as the Jahn-Teller and pseudo-Jahn-Teller effects~\cite{Kato2022}, which are relevant in the context of CISS. For example, in a recent work \cite{Das2025} a phonon-assisted spin-orbit coupling is proposed as the source of a topological anomalous Hall effect granting long-range spin-transport in chiral gold. In addition, edge and finite-size effects in p-orbital helical chains have been shown to generate nontrivial spin structures even with minimal SOC, highlighting the role of geometry and boundary conditions in spin  selection~\cite{Kato2023}.
Recently, Yao \textit{et al.}~\cite{Dapeng2025} formulated spin magnetization induced by chiral phonons in terms of the Berry curvature in phonon-displacement space, capturing the chiral nature of atomic vibrations. As remarked by these authors, an important aspect of spin generation by chiral phonons concerns dissipation and angular momentum conservation. Chiral phonons carry well-defined angular momentum associated with atomic rotations, which can be transferred to electronic spin and orbital degrees of freedom via spin–orbit coupling. \\
In a closed system, total angular momentum is conserved through exchange between lattice and electronic subsystems. However, realistic CISS experiments are inherently open and dissipative, involving coupling to leads, substrates, and phonon baths. In such cases, angular momentum can be redistributed to the environment, allowing for steady-state spin polarization under nonequilibrium conditions. Therefore, dissipation does not merely damp spin signals but plays a central role in stabilizing observable spin polarization driven by chiral vibrational modes. This complements earlier electronic and vibronic Berry-phase studies, extending geometric-phase concepts to both molecular and crystalline chiral systems. \\
\begin{figure*}
 \centering
 \includegraphics[height=8.5cm]{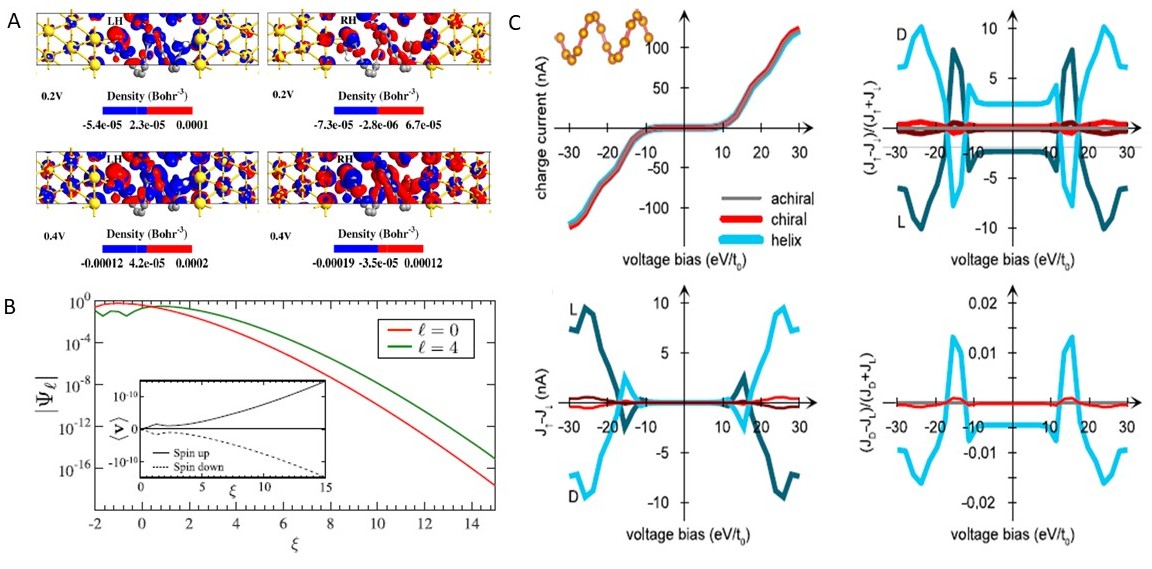}
 \caption{Illustrations of spin-dependent transport and polarization in chiral and helical systems. (A) Non-equilibrium spin density isosurfaces for left- and right-handed helices under different bias voltages with Au(111) electrodes. Reproduced from Ref. \citenum{Sumit2023} with permission. Copyright 2023, American Chemical Society. (B) Wavefunction tail amplitude $\xi \gg 1$ grows with angular momentum, with spin aligned to velocity. Reproduced from Ref. \citenum{Karen2019} with permission. Copyright 2019, American Chemical Society. (C) Transport properties for achiral (gray), chiral (red), and helical (blue) structures shown in Fig.~1. Here, the term ``helical" refers to a specific chiral geometry where $N=16$ sites are arranged along a spiral with azimuthal angle $\phi \in [0,4\pi]$ (inset in first panel). The darker blue curve represents the results for the helical configuration. Panels show charge current (upper left), current spin polarization (lower left), spin polarization normalized by the charge current (upper right), and enantiomeric asymmetry for a 20\% spin-polarized injected current (lower right). Here, the chiral enantiomers are labeled L (darker blue, helix) and D (blue, helix). The faint (in upper right panel) line indicates zero spin polarization. Reproduced from Ref. \citenum{Fransson2025} with permission. Copyright 2025, American Chemical Society.}
 \label{figure4}
\end{figure*}
In this vein, Subotnik and co-workers~\cite{Teh2022} demonstrated that nuclear wave packets propagating near conical intersections experience large Berry forces capable of inducing spin separation, even under weak SOC conditions. These findings imply that coupling between nuclear motion and electronic spin states, mediated by geometric phase effects, can generate spin polarization in light-atom molecules. 
Related work has shown that Berry curvature contributes an antisymmetric component to the electronic friction tensor that acts on nuclei embedded in a gapless electronic environment~\cite{Teh2021}, such as near metal surfaces. In reality, this antisymmetric term 
is spurious \cite{Martinazzo2022a,Martinazzo2022b}: 
in the common Markov limit of a fast electronic response, the 
overall effect of the electronic excitations (electronic friction) 
is to cool nuclear motion sufficiently to enforce a true adiabatic limit. The latter comes \emph{together with its gauge fields}, namely the Fubini–Study pseudo-electric and Berry pseudo-magnetic forces \cite{Martinazzo2022a,Martinazzo2022b,Martinazzo2023}. 
Hence, Berry forces can naturally emerge in electronically dissipative settings and may even account for equilibrium spin polarization in metallic systems. This reinforces the view that so-called “SOC-independent” routes to CISS are viable, provided that electronic excitations (non-adiabatic dynamics) are properly included.\\
While these studies explore spin dynamics arising from nuclear motion, the explicit role of chirality in such models is not always addressed. Nonetheless, quantum geometry offers additional promise. Recent work has applied the Fubini–Study metric to derive electron-phonon coupling strengths \textit{ab initio}~\cite{Yu2024}, revealing nontrivial contributions even in topologically trivial systems. These findings are especially relevant for CISS theories that rely on vibronic interactions and spin selectivity emerging from nuclear motion. \\ 
Another perspective emphasizes the role of orbital selectivity in CISS. In this view, chiral structures preferentially filter electronic states with specific orbital angular momentum, which can subsequently couple to spin degrees of freedom via spin–orbit interaction~\cite{Yoda2018, Liu2021}. Such orbital filtering mechanisms suggest that spin polarization may emerge from an initial orbital asymmetry, even when intrinsic atomic SOC is weak. Recent studies further explore how orbital texture, interface hybridization, and symmetry constraints contribute to spin–orbital conversion in chiral systems \cite{Adhikari2023,Yoda2018,Yang2025}.
\\
An intriguing but less explored direction involves the concept of curl forces, introduced by Berry and colleagues~\cite{Berry2015}. These nondissipative Newtonian forces, characterized by a non-zero curl, arise from kinetic energy terms that depend anisotropically on momentum. Since such forces cannot be derived from a scalar potential, they impose geometric constraints on particle trajectories and could potentially couple to spin degrees of freedom in chiral environments. We note that non-adiabatic dynamics generically generate non-conservative forces on the nuclei that act as electromotive forces, changing the (Berry) magnetic flux and thus the molecular geometric phase, in analogy with an \emph{inverse} Faraday–Maxwell induction law \cite{Martinazzo2024}. 
 Whether relativistic Hamiltonians for helical molecules naturally include these curl forces - and whether they discriminate between spin helicities - remains an open and promising question.
 Taken together, these diverse theoretical frameworks-ranging from topological band theory and Berry-phase dynamics to quantum geometric coupling and curl forces-offer mechanisms for spin polarization that are fundamentally distinct from traditional SOC-based explanations. \\
It is important to note that many existing theoretical treatments of CISS are formulated within effective single-particle frameworks. While such models capture essential symmetry and geometric aspects, they may overlook electron-electron interactions, correlation effects, and collective spin dynamics. Several recent works suggest that multiparticle mechanisms-such as interaction-enhanced spin polarization, spin fluctuations, and correlation-driven symmetry breaking—may play a crucial role in amplifying spin selectivity beyond single-particle expectations \cite{fransson25}. A comprehensive understanding of CISS likely requires combining symmetry-based approaches with many-body treatments. Future progress will likely require integrating symmetry-based single-particle frameworks with many-body treatments to fully resolve the magnitude and robustness of CISS.
\\
Although further work is needed to incorporate chirality explicitly in some of these models, their success in generating finite spin polarization under minimal SOC assumptions suggests that they may hold the key to a unified understanding of the CISS effect. These SOC-independent frameworks motivate a closer examination of dynamical spin processes in chiral systems. 

\section*{Spin Dynamics in CISS Modeling}
\begin{figure*}
 \centering
 \includegraphics[height=10.5cm]{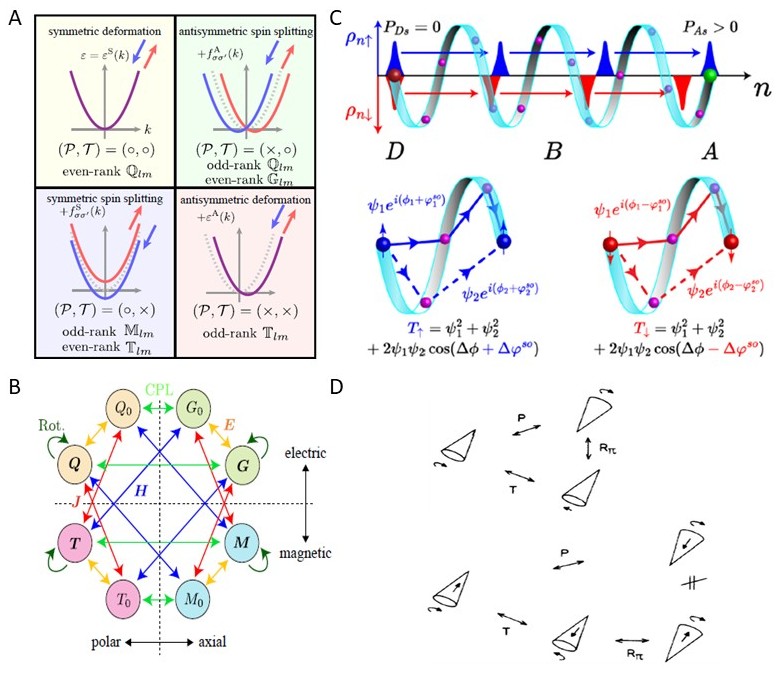}
 \caption{
 Schematics of spin-related effects under broken $P$ and $T$ symmetries, external fields, and rotational dynamics. 
 (A) 
 Splitting and shift between bands of different spin (indicated by the arrows) 
 with/without inversion ($P$) and time-reversal ($T$) symmetries. 
 (B) Relations between monopoles ($Q_0,M_0,T_0,G_0$) and dipoles ($Q,M,T,G$) via external fields (E, H, J, CPL, Rot.). Reproduced from Ref. \citenum{Hayami2024} with permission. Copyright 2024, Physical Society of Japan. (C) Schematics of spin-dependent velocities and separation induced by SOC and helicity, leading to transient spin polarization. Reproduced from Ref. \citenum{Zhang2025} with permission. Copyright 2025, American Physical Society. (D) 
 Spinning cone states illustrating the difference between true and false chirality as identified by their behavior under the combined effect of $P$ and $T$ transformations. Translating spinning objects and the concept of true chirality. A cone spinning about its symmetry axis appears chiral under space inversion $P$, but this chirality is false because the same configuration can be obtained by time reversal $T$ followed by a $180^\circ$ rotation about an axis perpendicular to the spin axis. When the spinning cone also translates along the spin axis, this equivalence breaks, resulting in true chirality. The arrow vectors indicate the directions of rotation (spin) and translation, which together determine the handedness. The same principle applies to translating spinning spheres.
 Reproduced from Ref. \citenum{Barron1986, Barron1986-2} with permission. Copyright 1986, American Chemical Society.}
 \label{figure5}
\end{figure*}

The emergence of spin-polarized currents in chiral systems, as described by the CISS effect, cannot be fully understood without considering the dynamical evolution of spin during charge transport \cite{Naaman2012, Brian2024}. Unlike spin selection in magnetically ordered systems, CISS relies on relativistic interactions-particularly SOC in concert with molecular chirality to induce spin asymmetry in otherwise nonmagnetic environments \cite{Dednam2023, Zhang2025} (a schematic representation is provided in Fig \ref{figure5} C). This spin selectivity arises not only from static electronic structure features, but also from time-dependent spin processes such as coherent precession, relaxation, and decoherence. These phenomena are sensitive to both intrinsic factors, such as molecular symmetry and SOC strength, and extrinsic conditions, including environmental interactions and thermal fluctuations \cite{Fransson2025}. In this section, we dissect the key aspects that govern spin dynamics in chiral media \cite{Hedegard2023}. First, we 
discuss the mechanisms of spin relaxation and decoherence that influence the preservation of spin polarization during transport \cite{Jan2025}. We then add how spin-dependent scattering and interference give rise to dynamic spin filtering in the presence of SOC \cite{Karen2019}. Finally, we highlight current computational strategies used to model spin evolution in chiral systems, encompassing both quantum-coherent and open-system approaches. Together, these perspectives provide a comprehensive understanding of how spin dynamics underpins spin selectivity in CISS-active materials (Fig. \ref{figure4}). \\
The low symmetry and structural complexity of chiral molecules often lead to anisotropic and spatially varying SOC fields, which modulate spin relaxation pathways in nontrivial ways. A key condition for effective spin filtering in the CISS framework is the alignment of spin coherence time with the electron’s dwell time in the chiral region. If the spin decoheres too quickly, SOC-driven filtering mechanisms are disrupted; if coherence is maintained sufficiently, spin asymmetry can be significantly amplified. Theoretical approaches such as Lindblad master equations and open-system density matrix evolution have been instrumental in clarifying how coherence loss affects spin selectivity \cite{Francesco2024}. These studies suggest that CISS operates in a delicate regime where partial coherence and controlled decoherence are both essential to the observed spin response \cite{Naaman2019}. \\
The transport of spin through chiral molecules is inherently time-dependent and governed by non-equilibrium processes. Unlike conventional spin transport mechanisms, CISS does not rely on intrinsic magnetic order but instead emerges from the entanglement of spin and spatial degrees of freedom 
in the presence of SOC and structural chirality \cite{Krieger2024}. As charge carriers move through a chiral potential landscape, SOC induces a spin-dependent asymmetry in transmission probabilities-favoring spins aligned with the handedness of the molecular system \cite{Gohler2011}. Among the reported experimental benchmarks, particularly large spin polarization values were achieved by Al-Bustami \textit{et al.}~\cite{Al-Bustami2022}, representing among the highest CISS-induced spin selectivities reported to date. These results are in a dynamic form of spin filtering where spin selectivity emerges during the transport process itself. The efficiency of this filtering is contingent on the interplay between several time scales: the spin precession frequency (determined by SOC), the electron’s dwell time in the chiral medium. Only when these scales are favorably matched does dynamic spin polarization manifest robustly. Environmental interactions, such as stochastic fluctuations or vibronic coupling, can either enhance or suppress this effect depending on their strength and temporal structure. Modeling approaches that explicitly incorporate these 
dynamical effects - such as time-dependent NEGFs (TD-NEGF) formalism,
high-dimensional wave-packet propagation, and stochastic Schrödinger equations - have shown that even weak SOC can give rise to significant spin polarization when combined with molecular asymmetry and partial coherence \cite{Sumit2023} (see Fig. \ref{figure4} A).  
Thus, the CISS effect reflects a fundamentally dynamical mechanism where spin filtering evolves over time under the influence of both coherent and dissipative interactions \cite{Koyel2018}. \\ 
Capturing the full complexity of spin dynamics in CISS systems requires computational methods that account for both relativistic effects and open-system behavior. Time-dependent DFT, particularly in its four-component relativistic form, provides an \textit{ab initio} framework for describing SOC and chiral electronic structure \cite{Daniel2020}. However, TDDFT alone is often insufficient for modeling non-equilibrium charge transport or spin decoherence, which are central to the dynamical nature of CISS \cite{Fransson2023, Francesco2024}. To address this, NEGF-based transport simulations have been extended to include spin degrees of freedom and SOC, allowing for the computation of spin-resolved currents through molecular junctions under bias \cite{Hedegard2023}. Complementary to these are open quantum system approaches-including Lindblad formalisms \cite{Francesco2024}, Redfield theory \cite{Thomas2021}, and stochastic methods-which model~\cite{Bloch1953} the interaction of spins with thermal and vibrational baths, thereby capturing relaxation and dephasing dynamics. On a mesoscopic level, tight-binding Hamiltonians augmented with effective SOC terms provide a versatile platform for exploring how symmetry, energy level alignment, and disorder influence spin transport. Recent developments have also integrated machine learning with quantum dynamics simulations to efficiently explore vast molecular design spaces and predict spin-filtering behavior across diverse chiral scaffolds \cite{Kato2025}. These computational advancements, when combined, enable a multiscale perspective on spin dynamics, bridging fundamental relativistic interactions with observable spin-selective transport phenomena \cite{Liang2025}. To further rationalize these effects, symmetry-based and field-theoretical approaches provide a unifying understanding and perspective.

\section*{Integrating Field and Group Theoretical Frameworks}
A comprehensive understanding of the CISS effect necessitates the integration of relativistic field-theoretical formulations with symmetry-based group-theoretical frameworks \cite{Bencheikh2003}.~The Dirac equation~\cite{Daniel2020}, as a first-order relativistic wave equation, inherently incorporates the coupling between spin and linear momentum through its spinor structure \cite{Truhlar2025}. Within this formalism, the $\gamma^5$ matrix~\cite{musielak2021} serves as the generator of chirality, allowing the decomposition of Dirac spinors into left- and right-handed components \cite{Su2023}. \\
In the weak-relativistic limit, the expectation value of $\gamma_5$ reduces to the helicity operator -- \textit{i.e.}, the projection of the electron spin onto its momentum direction -- thus acting as a pseudoscalar quantity that is $\mathcal{P}$-odd and $\mathcal{T}$-even. This symmetry behavior corresponds precisely to the definition of true chirality as introduced by Barron~\cite{Barron1986} (see Fig. \ref{figure5} D, from the same reference, for a visual comparison between true and false chirality, based on spinning cones) and later 
generalized by M. Petitjean who extended the definition of chirality to non-Euclidean metric spaces and to Minkowski's space-time  \cite{petitjean19,petitjean20,petitjean20_1,petitjean22arxiv}, notably connecting to particles chirality as defined through Dirac's $\gamma$ matrices. Another interesting view of chirality was achieved through its direct connection to the concept of
the electric toroidal monopole ($G_0$) within the complete multipole representation (CMR) formalism (see Fig. \ref{figure5} A, and B). The CMR framework provides a systematic basis for classifying all possible symmetry-adapted electronic degrees of freedom-including charge, spin, orbital angular momentum, and their combinations via a complete set of multipole operators. Crucially, it extends the conventional electric ($\mathcal{T}$-even, $\mathcal{P}$-even) and magnetic ($\mathcal{T}$-odd, $\mathcal{P}$-odd) multipoles by including electric toroidal ($\mathcal{T}$-even, $\mathcal{P}$-odd) and magnetic toroidal ($\mathcal{T}$-odd, $\mathcal{P}$-even) components, which are indispensable for representing chiral and time-reversal symmetry-breaking phenomena \cite{musielak2021}. 
Within this framework, the $G_0$ monopole plays a central role in describing the microscopic sources of chirality and has been explicitly linked to the helicity of Dirac spinors in relativistic quantum mechanics. This correspondence establishes a rigorous bridge between field-theoretical constructs and group-theoretical symmetry classifications. 
Moreover, it is worth noting the growing interest in the non-Hermitian skin effect \cite{Yao2018,Yokomizo2019,Zhang2020_1,Yi2020,Okuma2023}, which, by breaking the conventional bulk–edge correspondence, can lead to the localization of bulk eigenstates at the system boundaries (i.e., under open boundary conditions).  
This phenomenon has already been invoked to explain violations of Onsager reciprocity in chiral systems, in conjunction with charge trapping effects, through a magnetochiral charge pumping mechanism \cite{Zhao2025}. \\
%
By leveraging this dual formalism, it becomes possible to identify and characterize the active multipoles that transform according to irreducible representations of the system's (magnetic) point group and contribute to spin-polarized transport. External perturbations such as electric fields, magnetic fields, or charge currents can couple to internal degrees of freedom through multipole interconversion, leading to the activation of chiral responses \cite{Hirakida2025, Li2020}. In particular, terms such as the relativistic $\mathbf{p} \cdot \boldsymbol{\sigma}$ helicity operator, which appears in transport models of CISS, can be formally understood as symmetry-allowed couplings between specific $G$-type and $M$-type multipoles in the CMR scheme. Furthermore, spin-resolved response tensors, including spin conductivity and magnetoelectric coupling, can be systematically constructed using the multipole basis, allowing one to trace the contributions of individual symmetry channels. \\
The integration of relativistic quantum chemistry (e.g., four-component DFT) with CMR analysis offers a promising path forward. It enables the extraction and quantification of multipole moments associated with specific spin and orbital configurations, potentially overcoming the limitations of conventional (non-relativistic) DFT approaches that fail to capture the subtle spin–chirality interplay inherent in CISS. This integrated approach may also facilitate the construction of functionals based on toroidal multipole densities in extended DFT frameworks, or the derivation of symmetry-constrained asymmetric spin–orbit coupling (ASOC) terms from first principles \cite{Zhou2024}. Altogether, this synthesis of field-theoretical and group-theoretical perspectives~\cite{Li2020} provides a robust platform for decoding the microscopic origin of spin selectivity in chiral systems \cite{Zhou2024} and developing predictive models for spin-dependent transport phenomena \cite{Hoshino2023}. \\
Despite significant progress, several fundamental theoretical questions in CISS remain actively debated. These controversies 
include (but are not limited to) the magnitude of spin–orbit coupling required, the adequacy of single-particle descriptions and the importance of interactions (e.g., electron-electron or electron-phonon), the role of time-reversal symmetry, the influence of dissipation and many-body interactions, the role of the leads and their interface with the system in transport experiments, and finally the connection with topological states/phases. 
Addressing these challenges will require coordinated advances in theory, materials design, and precision experiments. A unified understanding of CISS is likely to emerge from integrating symmetry-based insights with realistic many-body and non-equilibrium approaches.

%
%
%
%
%
%
%
%
%
%

\section*{Conclusions and Future Prospects}
This mini-review surveys the emergence of the CISS effect from both experimental and theoretical standpoints, focusing on SOC and spin dynamics in light-element chiral systems. Despite the absence of a unified theoretical framework, two complementary approaches—relativistic quantum mechanics and complete multipole analysis—offer a promising path forward by connecting spin–momentum coupling with chirality through symmetry principles. Computational evidence supports a direct link between molecular structure and spin polarization, notably revealing a correspondence between chirality density and spin current in the weak-relativistic limit. Moreover, external fields introduce additional couplings that enrich the theoretical landscape, establishing a solid basis for microscopic models and predictive simulations of spin-selective transport.  

Looking ahead, progress in unraveling spin selectivity will rely on integrating phenomenological models with predictive \textit{ab initio} approaches. Advancing current- and spin-current DFT, together with relativistic field-theoretical formulations (e.g., based on the use of covariant derivatives including the effect of possibly non-abelian gauge fields) of curvature- and torsion-induced SOC, will be essential to quantify emergent interactions beyond atomic limits. Multiscale simulations that combine relativistic quantum chemistry with open-system dynamics (NEGF, TDDFT, and Lindblad formalisms) are needed to capture coherence, decoherence, and vibronic couplings. Symmetry-based multipole expansions provide a rigorous framework to classify spin–chirality couplings across molecules, interfaces, and crystals. On the experimental side, engineered chiral 2D systems and hybrid organic–inorganic platforms will serve as key testbeds, while machine learning-assisted screening can accelerate the discovery of molecular scaffolds with enhanced spin polarization. Together, these directions promise a quantitative, symmetry-based foundation for the rational design of chiral platforms in spintronics, quantum information, and catalysis.

\section*{Appendix: Spin density and (transport) spin polarization}

For the sake of clarity, in this Appendix we focus on the somewhat ambiguous use of the term ``spin polarization” and on the more precise meaning adopted in the main text. Spin polarization always refers to the spin state of “something,” but it is not always clear what this “something” is. In the main text, we specifically used it to describe the spin state of the transmitted current, while the terms spin density or magnetization were reserved for characterizing electron distributions.

For simplicity we focus on a single electron, but the arguments given here are easily generalized to many-electron systems.
The spin polarization of an electron distribution described by the spinor $\psi(\mathbf{r}) \equiv \{\psi_{\sigma}(\mathbf{r})\}$
is simply the spin-density or (spin) magnetization associated to the distribution
\begin{equation}
\mathbf{m}(\mathbf{r}) = \psi^{\dagger}(\mathbf{r})\,\boldsymbol{\sigma}\,\psi(\mathbf{r})
= \mathrm{tr}_s\!\left(\boldsymbol{\sigma}\rho^{s}(\mathbf{r})\right)
\end{equation}
where $\mathrm{tr}_s$ denotes the trace over the spin degrees of freedom,
$\rho^{s}(\mathbf{r})=\psi(\mathbf{r})\psi^{\dagger}(\mathbf{r})$ is the spin density matrix at position $\mathbf{r}$ and $\boldsymbol{\sigma}$ is the vector of Pauli matrices representing the spin operator in the standard basis, for some choice of the reference system
(here, $^{\dagger}$ is the ordinary Hermitian conjugation in $\mathbb{C}^{2}$).
This is the spin-vector component of the local spin-density matrix $\rho^{s}(\mathbf{r})$ in the spin decomposition
\begin{equation}
\rho^{s}(\mathbf{r}) = \frac{1}{2}\left[\rho(\mathbf{r}) + \mathbf{m}(\mathbf{r})\!\cdot\!\boldsymbol{\sigma}\right]
\end{equation}
where $\rho(\mathbf{r})$ is the ordinary (number) density.
The latter should be considered jointly with the (number) density current
\begin{equation}
\mathbf{j}(\mathbf{r}) =
\Re\!\left[\psi^{\dagger}(\mathbf{r})\,\mathbf{v}\,\psi(\mathbf{r})\right]
= \Re\!\left[\mathbf{v}\,\rho(\mathbf{r},\mathbf{r}’)\right]\Big|_{\mathbf{r}’=\mathbf{r}}
\end{equation}
where $\mathbf{v}=\boldsymbol{\pi}/m$ is the velocity operator, $\boldsymbol{\pi}=\mathbf{p}-q\mathbf{A}/c$ being the mechanical momentum,
with $q$ the particle charge, $c$ the speed of light, and $m$ the mass.
In the second step, the notation $\mathbf{r}’=\mathbf{r}$ means that $\mathbf{r}’$ must be set equal to $\mathbf{r}$ after the velocity operator has acted on the $\mathbf{r}$-dependence of $\rho$.

Similarly to the current density, one can introduce the current spin density
\begin{equation}
j_{\alpha k}^{s}(\mathbf{r})
= \Re\!\left[\psi^{\dagger}(\mathbf{r})\,\sigma_{\alpha}\,v_{k}\,\psi(\mathbf{r})\right]
= \Re\!\left[\sigma_{\alpha}v_{k}\rho(\mathbf{r},\mathbf{r}’)\right]\Big|_{\mathbf{r}’=\mathbf{r}}
\end{equation}
This is a rank-2 tensor whose $(\alpha,k)$ component gives the $\alpha$-component of the spin current in the k-direction.
Greek and Latin indices are used here to emphasize the different meaning of the two Cartesian coordinates, although they transform in the same way under ordinary SO(3) rotations. When the particle current has a definite direction $\hat{\mathbf{n}}$, we can focus on this direction and define the spin-current \emph{vector}
\begin{equation}
\mathbf{j}^{s}(\mathbf{r})
= \Re\left[\psi^{\dagger}(\mathbf{r})\,\boldsymbol{\sigma}\,v_{n}\,\psi(\mathbf{r})\right]
= \Re\left[\boldsymbol{\sigma}v_{n}\rho(\mathbf{r},\mathbf{r}’)\right]\Big\vert_{\mathbf{r}’=\mathbf{r}}
\end{equation}
where $v{n}=\mathbf{v}\!\cdot\!\hat{\mathbf{n}}$ denotes the component along $\hat{\mathbf{n}}$.
This is a vector tied to the current flow, connected to the spin state of the moving electron. If we want to characterize the spin state of the moving electron, we need to introduce a “velocity-weighted” spin-density matrix
\begin{equation}
\rho_{v}^{s}
=\frac{(v_{n}\psi)(\mathbf{r})\psi^{\dagger}(\mathbf{r})+\psi(\mathbf{r})\big(v_{n}\psi\big)^{\dagger}(\mathbf{r})}{2j},
\qquad
j=\Re\!\left[\psi^{\dagger}v_{n}\psi\right]
\end{equation}

This matrix is Hermitian and correctly normalized with respect to the magnitude of the number current density j (provided j>0).
It is also positive definite in the quasi-collinear approximation, that is, when the spinor $\psi$ and its velocity-weighted counterpart $v_{n}\psi$ are nearly collinear in spin space — a condition usually satisfied for well-defined transport states.
The spin polarization of the moving electron is then
\begin{equation}
\mathbf{P}^{s}
= \mathrm{tr}_{s}\!\left(\boldsymbol{\sigma}\rho_{v}^{s}\right)
= \frac{\mathbf{j}^{s}}{j}
\end{equation}
This definition suffices for our scopes, though it fails in describing pure spin currents, where $\mathbf{j}^{s}\neq\mathbf{0}$ despite $j=0$.
In such cases the ratio $\mathbf{j}^{s}/j$ is ill-defined, and the spin-current tensor $\mathbf{j}^{s}$ itself — rather than any normalized polarization — is the meaningful quantity.

We notice that the spin current density connects to the changes in time of the spin-density in a way similar to how the ordinary current density connects to charge variations,
\begin{equation}
\partial_{t}\rho = -\boldsymbol{\nabla}\!\cdot\!\mathbf{j} = -\sum_k \partial_k j_k
\end{equation}
Consider, for instance, a Pauli–Schrödinger Hamiltonian
\begin{equation}
H = \frac{(\boldsymbol{\pi}\boldsymbol{\sigma})^{2}}{2m} + V
\end{equation}
and the time evolution of the magnetization
\begin{equation}
\partial_{t}\mathbf{m}(\mathbf{r}) = \frac{2}{\hslash}\,\Im\!\left(\psi^{\dagger}\boldsymbol{\sigma}H\psi\right)
\end{equation}
Using the identity $(\boldsymbol{\pi}\times\boldsymbol{\pi})=i\hslash q\mathbf{B}/c$, which introduces the magnetic field $\mathbf{B}$, one finds
\begin{equation}
\boldsymbol{\sigma}(\boldsymbol{\sigma}\!\cdot\!\boldsymbol{\pi})^{2}
= \boldsymbol{\sigma}\pi^{2} - \hslash\frac{q}{c}\mathbf{B} +	i\hslash\frac{q}{c}\boldsymbol{\sigma}\times\mathbf{B}
\end{equation}
and therefore
\begin{equation}
\partial_{t}\mathbf{m}(\mathbf{r})
= -\sum_{k}\partial_{k}(\mathbf{j}^{s})_{k} + \frac{q}{mc}\,\mathbf{m}\times\mathbf{B}
\end{equation}
where $(\mathbf{j}^{s})_{k}$ is the $k^\textrm{th}$ column of the spin-current tensor. Clearly, the first term on the right-hand side describes the change of the spin density due to electron dynamics, while the second describes spin precession around the magnetic field (if any).
Additional terms in the Hamiltonian (e.g., a spin–orbit coupling term of the form $\boldsymbol{\Lambda}\!\cdot\!\boldsymbol{\sigma}$, involving the momentum operator in the vector $\boldsymbol{\Lambda}$) introduce further torque contributions on the right-hand side, but do not alter the meaning of $\mathbf{j}^{s}$.

Therefore, one should distinguish between a finite spin density in equilibrium and the notion of spin polarization in transport. The former reflects a static property of the electronic ground state, while the latter refers to the spin character of the current collected at a drain terminal.

\section*{Conflicts of interest}
There are no conflicts to declare.

\section*{Data availability}
This manuscript is a mini-review and does not include, generate, or analyze primary research data, software, or code. All information discussed is derived from previously published journal articles and PhD/Master theses.

\section*{Acknowledgements}
This work is supported by the PRIN initiative of the Italian Ministry of Research and University (MUR), under grant no. F53D23001070006-CISS-PRIN22-2022FL4NZ4. 

\bibliography{rsc-rev}

\end{document}